\documentclass[
preprint,
aps,
prd,
superscriptaddress,
nofootinbib,
amsmath
]{revtex4}
\usepackage{bm}
\usepackage[dvips]{color,graphicx}
\usepackage{epsfig}
\usepackage[figuresright]{rotating}
\allowdisplaybreaks[4]
\newcommand{\ud}{\mathrm{d}}
\newcommand{\eq}[1]{Eq.~(\ref{#1})}

\newcommand{\eqs}[1]{Eqs.~(\ref{#1})}

\newcommand{\Tr}{\mathop{\mathrm{Tr}}}
\newcommand{\tr}{\mathop{\mathrm{tr}}}
\newcommand{\ceps}{\varepsilon}
\newcommand{\eps}{\varepsilon_{\mu\nu\alpha\beta}}
\newcommand{\sop}{\widehat{\mathcal{S}}}

\newcommand{\intp}{\int[\ud p]}
\newcommand{\spfn}[2]{\mathcal{P}^{#1}(\varepsilon,\bm{p},{#2})}
\begin{document}

\title{Deuteron Spin Structure Functions \\
	in the Resonance and DIS Regions}

\author{S. A. Kulagin}	
\affiliation{Institute for Nuclear Research,
	Russian Academy of Sciences,
	117312 Moscow, Russia}

\author{W. Melnitchouk}	
\affiliation{Jefferson Lab,
	12000 Jefferson Avenue,
	Newport News, VA 23606, USA}

\begin{abstract}
We derive relations between spin-dependent nuclear and nucleon $g_1$
and $g_2$ structure functions within the nuclear impulse approximation,
which are valid at all $Q^2$, and in both the resonance and deep
inelastic regions.
We apply the formalism to the specific case of the deuteron, which is
often used as a source of neutron structure information, and compare
the size of the nuclear corrections calculated using exact kinematics
and using approximations applicable at large $Q^2$.
\end{abstract}

\maketitle

{\bf 1. Introduction}
\vspace*{0.5cm}

The study of nuclear effects in deep inelastic structure functions
has by now a long and rich history.
The importance of nuclear structure in high-energy scattering of
leptons from nuclei was most prominently thrust into the limelight
by the ``nuclear EMC effect'' discovered by the European Muon
Collaboration (EMC) \cite{EMC}, which found much larger
than anticipated differences between structure functions of heavy
nuclei and those of deuterium.
Many theoretical and experimental studies of nuclear effects on
structure functions followed, and over the years an extensive
phenomenology has been developed, even though many questions about
the origin of the effect still remain (for reviews see {\em e.g.}
Ref.~\cite{EMCrev}).

While the early studies of nuclear medium modifications focussed on
heavy nuclei, where the magnitude of the effects is largest, it has
since been realized that resolving the dynamical origin of the EMC
effect requires understanding of light nuclei as well.
Until recently, an anomalous situation existed whereby nuclei with
$A \lesssim 4$, for which the most detailed microscopic theoretical
calculations were possible, had the least empirical information
available.
A dedicated experiment at Jefferson Lab \cite{E03-103}, with the goal
of making precise measurements of the nuclear dependence of structure
functions in a variety of light nuclei, has recently completed taking
data which are currently being analyzed.

Light nuclei, such as deuterium or $^3$He, are also often used as
effective neutron targets in experiments seeking to extract information
on the structure of free neutrons.
This has been especially relevant for neutron spin structure studies,
for which polarized $^3$He is commonly used.
If one is to extract reliable information on the structure of the
neutron, it is important that the nuclear corrections are properly
accounted for, especially given the ever increasing precision of modern
experiments.

In recent years several theoretical analyses have been devoted to the
nuclear EMC effect in $A=2$ and $A=3$ nuclei, both for unpolarized [4-16]
and polarized [17-25]
structure functions,
which have quantified the effects of binding, Fermi motion, as well
as relativity and nucleon off-shell effects.
Most of these studies have focussed on nuclear effects in the deep
inelastic scattering (DIS) region, where the exchanged photon's
virtuality $Q^2$ and the mass $W$ of the final hadronic state are
both large ($Q^2, W^2 \gg 1$~GeV$^2$).

On the other hand, considerable data have been accumulated recently
for $W \lesssim 2$~GeV, in the region dominated by nucleon resonances.
The resonance region has received renewed interest partly due to the
remarkable phenomenon of Bloom-Gilman duality \cite{BG}, in which
averages of resonance structure functions have been found to
approximately equal the scaling region functions, measured at much
larger $W$ (see Ref.~\cite{MEK} and references therein).
Extracting information on the neutron in the resonance region from
nuclear data is particularly challenging.

The resonance region has considerably richer structure because of
the prominence of specific resonances in the inclusive spectrum,
such as the $\Delta(1232)$ or the $S_{11}\ N^*(1535)$ resonances.
As well as exhibiting pronounced peaks, these resonance structures
are also strongly $Q^2$ dependent.
At high $Q^2$ the role of resonances is restricted by kinematics
to the $x \sim 1$ domain, however, at low $Q^2$ ($\sim 1$~GeV$^2$)
resonances actually dominate the cross section.
The effects of nucleon Fermi motion are expected to wash out much
of these structures in a nucleus, leading to dramatic differences
between structure functions of nucleons and nuclei, and hence much
more interesting EMC effects than in the deep inelastic region.

The theoretical tools for the study of nuclear effects in structure
functions have, for historical reasons, been developed specifically
for the DIS region, usually assuming the Bjorken limit
(where both $Q^2$ and the energy transfer to the target $\to \infty$),
in which the kinematics greatly simplify.
To date, however, only approximate methods have been used to describe
nuclear corrections in the resonance and low-$Q^2$ regions, mostly
using either effective polarizations, or convolution approximations
which are valid strictly only in the Bjorken limit.

In this paper we rectify this problem.
Working within the nuclear impulse approximation, in which the virtual
photon interacts with a single nucleon inside the nucleus, we derive a
set of formulas for the spin-dependent nuclear $g_1$ and $g_2$ structure
functions, which are valid over a broad range of kinematics.
In fact, since the derivation involves an exact treatment of finite
$Q^2$ kinematics, and does not depend on the twist expansion,
our results are valid for any $Q^2$, and in both the deep inelastic and
resonance regions (and even for real photoproduction).

The general derivation is made within a relativistic framework; however,
for practical applications we specialize to the weak binding limit (WBL),
in which the nuclear matrix elements are systematically expanded in
powers of nucleon momentum $\bm{p}/M$, where $M$ is the nucleon mass.
In this limit, we find that the nuclear $g_1^A$ and $g_2^A$ structure
functions can be written as generalized convolution of the nucleon
$g_{1,2}^\tau$ structure functions ($\tau = p, n$) and spin-dependent
nucleon energy--momentum distributions in nuclei.
At finite $Q^2$, the $g_{1,2}^A$ functions receive contributions from
both $g_1^\tau$ and $g_2^\tau$, in contrast to the Bjorken limit results
which are diagonal for $g_1$ \cite{SS,Italy,ScopHe3}.

Our formal results are quite general and applicable to all nuclei.
However, in view of the current importance of understanding of nuclear
corrections in light nuclei, we demonstrate our formalism by applying
it to the simplest nucleus, namely the deuteron.
In Sec.~2 we outline the derivation of the nuclear structure functions
in the weak binding limit, and present for the first time the complete
set of formulas for polarized nuclear structure functions in terms of
those of bound nucleons, valid at arbitrary $Q^2$.
At finite $Q^2$ we find a breakdown of the simple convolution
expressions for the nuclear structure functions, in which all of the
$Q^2$ dependence is absorbed into the nucleon structure functions,
with the nucleon momentum distributions being functions of the nucleon
momentum and energy only.
In the generalized convolution that we derive at finite $Q^2$, the
nucleon momentum distributions depend in addition on the parameter
$\gamma \equiv |\bm{q}|/q_0$, where $\bm{q}$ and $q_0$ are
momentum and energy transfer, respectively.

In Sec.~3 we apply the formalism to the specific case of the deuteron,
and study the dependence of the finite-$Q^2$ nucleon momentum
distributions on the parameter $\gamma$.
The numerical results for the deuteron structure functions are presented
in Sec.~4, where we focus in particular on the EMC effect in the nucleon
resonance region, and contrast this with the effect for leading twist
structure functions.
Finally in Sec.~5 we summarize our results and discuss future work.

\vspace*{1cm}
{\bf 2. Nuclear structure functions in the weak binding limit}
\vspace*{0.5cm}

We begin our discussion of the nuclear structure function by
summarizing the results within the framework of the relativistic
nuclear impulse approximation, in which the scattering from the
nucleus proceeds via the scattering from its individual nucleon
constituents.  Corrections to the impulse approximation, in the form
of multiple scattering from nucleons, or meson-exchange currents,
are typically confined to the small-$x$ region, and can be safely
neglected by restricting the analysis to $x \agt 0.2$. 
Possible corrections due to the final state interaction of the spectator
nucleon with the hadronic debris are also not considered here.

\vspace*{1cm}
{\em 2 a. Hadronic tensor}
\vspace*{0.5cm}

In the impulse approximation the hadronic tensor for a nucleus with
four-momentum $P_A$ and spin $S$ can be expressed in terms of the
nucleon propagator in a nucleus and the virtual photon Compton
amplitude for a bound nucleon (for more details see {\em e.g.}
Refs.~\cite{MST,KPW,KP,MPT,KMPW}):
\begin{equation}
\label{WA}
W_{\mu\nu}^A(P_A,q,S) =
\int [\ud p]\, \Tr\left[
{\cal A}^\tau(p,P_A,S)\ {\cal W}_{\mu\nu}^\tau(p,q)
\right],
\end{equation}
where the sum is taken over bound protons and neutrons ($\tau=p,n$),
and the integration is performed over the nucleon four-momentum $p$,
for which we use the contracted notation $[\ud p]=\ud^4p/(2\pi)^4$.
In \eq{WA}, ${\cal A}^\tau(p,P_A,S)$ is the imaginary part of the
proton ($\tau=p$) or neutron ($\tau=n$) propagator in a nucleus $A$
with momentum $P_A$ and spin $S$: 
\begin{equation}
\label{A}
{\cal A}_{\alpha\beta}^\tau(p,P_A,S)=
\int \ud t\ \ud^3\bm{r}\ e^{i (p_0 t - \bm{p}\cdot\bm{r})}\
\langle P_A,S|\
	\overline{\Psi}_\beta^\tau(t,\bm{r})\,\Psi_\alpha^\tau(0)\
| P_A,S \rangle\ ,
\end{equation}
with $\Psi_\alpha^\tau(t,\bm{r})$ the (relativistic) nucleon field
operator, and $\alpha$ and $\beta$ Dirac spinor indeces.
The bound (off-shell) nucleon electromagnetic tensor
${\cal W}^\tau_{\mu\nu}(p,q)$ in \eq{WA} is a matrix in Dirac space,
and the trace ``Tr'' is taken in the nucleon Dirac space.

The analysis of \eq{WA} in the fully relativistic case requires
the solution to the nuclear bound state problem [in particular the
calculation of ${\cal A}^\tau(p,P_A,S)$], a task yet to be fully
completed.%
\footnote{We refer in this context to calculations based on
Bethe-Salpeter and light-cone approaches \cite{U:1996,Karmanov}.}
The presence of nucleon spin introduces some complications in \eq{WA},
such as additional Lorentz--Dirac structures (structure functions)
in the hadronic tensor ${\cal W}^\tau_{\mu\nu}(p,q)$ for the bound
nucleon \cite{MST,KPW}.
Nevertheless, the analysis can be significantly simplified in the target
rest frame within a nonrelativistic approximation, or {\em weak binding
limit} (WBL), assuming typical nucleon momenta and energies to be small
compared to the nucleon mass.

Starting from the general expression for ${\cal W}^\tau_{\mu\nu}(p,q)$
for the off-shell nucleon, and performing a systematic expansion of the
matrix elements in \eq{WA} in terms of $|\bm{p}|/M$ and $\ceps/M$ up to
order $\bm{p}^2/M^2$ and $\ceps/M$, where $\ceps=p_0-M$ is the nucleon
separation energy, one can show \cite{KPW,KP,KMPW} that the nuclear
hadronic tensor can be written in terms of the (nonrelativistic) nucleon
spectral function in the nucleus, $\mathcal{P}^\tau(\ceps,\bm{p},S)$,
and the bound nucleon hadronic tensor,
\begin{equation}
\label{WA:NR}
\frac{1}{M_A} W_{\mu\nu}^A(P_A,q,S) =
\int [\ud p]\ \frac{1}{M+\varepsilon} \tr\ [\spfn{\tau}{S}\,
\mathcal{W}_{\mu\nu}^\tau(p,q)]\ .
\end{equation}
Here $M_A$ is the nuclear mass, and the trace ``tr'' is taken over
nucleon polarization states.

The spin-dependent part of the nuclear hadronic tensor is related
to the antisymmetric part of the off-shell nucleon tensor
$\mathcal{W}_{\mu\nu}^\tau$, which depends on the nucleon polarization.
In the WBL this component of the hadronic tensor of the nonrelativistic
bound nucleon is given by:
\begin{equation}
\label{eq:W:NR}
\mathcal{W}^\tau_{\mu\nu}(p,q)
= \frac{M}{p\cdot q}
\left[
  (g_1^\tau + g_2^\tau)\, \eps\ q^\alpha \sop^\beta 
- g_2^\tau\, \eps\ q^\alpha p^\beta \frac{\sop\cdot q}{p\cdot q}
\right]\ ,
\end{equation}
where $g_{1,2}^\tau$ are the spin structure functions of the off-shell
proton or neutron ($\tau=p,n$) with four-momentum $p=(M+\ceps,\bm p)$.
The operator $\sop$ in Eq.~(\ref{eq:W:NR}) has a structure similar to
that of the spin four-vector $(0,\bm{\sigma})$ boosted to a reference
frame in which the nucleon has (nonrelativistic) momentum $\bm{p}$:
\begin{equation}
\label{eq:S:NR}
\sop = \left(\frac{\bm{\sigma}\cdot\bm{p}}{M},\,
\bm{\sigma}+\frac{\bm{p}\,(\bm{\sigma}\cdot\bm{p})}{2M^2}\right)\ .
\end{equation}

In Eq.~(\ref{WA:NR}) the nuclear hadronic tensor $W_{\mu\nu}^A$
factorizes into high-energy (${\cal W}_{\mu\nu}$) and low-energy
(${\cal P}$) domains in the nuclear scattering amplitudes.
The presence of spin, however, implies that this factorization does
not in general carry over to a corresponding factorization of nuclear
structure functions \cite{MST,KPW,MPT,KMPW}, unless further assumptions
or simplifications are made.

The relation between the nuclear and the nucleon spin structure
functions for polarized $^3$He and for the deuteron was discussed
in Refs.~\cite{SS,Italy,ScopD,ScopHe3}.
Note that the implementation of the impulse approximation is
different in Ref.~\cite{SS} and in the present approach
(see also Ref.~\cite{KMPW}).
Here we begin from a relativistic impulse approximation, \eq{WA},
which fully takes into account the nucleon spin degrees of freedom,
and work within the WBL to obtain the nuclear hadronic tensor in
terms of the nuclear spectral function, \eq{WA:NR}.
The latter equation is similar to the corresponding relation in
Ref.~\cite{SS}, but is not identical due to a different treatment
of the reaction kinematics.
In particular, in Ref.~\cite{SS} the struck nucleon is assumed to be
on the mass shell, while here we take the nucleon to be off-mass-shell.
Further details of the derivation of \eq{WA:NR} will be given elsewhere
\cite{KM}.

\vspace*{1cm}
{\em 2 b. Spectral function}
\vspace*{0.5cm}

The nuclear spectral function ${\mathcal P}$ is defined similarly
to the nucleon propagator in the nucleus ${\cal A}$ in \eq{A},
but involves the correlator of nonrelativistic two-component nucleon 
operators: 
\begin{equation}
\label{eq:spfn}
{\mathcal P}^\tau_{\sigma\sigma'}(\varepsilon,\bm{p},\bm{S}) =
\int \ud t\,\ud^3r\, \ud^3r' \,
e^{i (\varepsilon t-\bm{p}\cdot(\bm{r}'-\bm{r}))}
\left\langle
{\psi^\tau_{\sigma'}}^\dagger(\bm{r}',t)\ \psi_{\sigma}^\tau(\bm{r},0)
\right\rangle\ ,
\end{equation}
where the nucleon operator $\psi_{\sigma}^\tau$ describes the nucleon
with polarization $\sigma$ and isospin $\tau$,
and the average is taken
over the nuclear ground state.
The nuclear spin is described in \eq{eq:spfn} by the axial three-vector
$\bm{S}$. 
For spin-1/2 targets, $\bm{S}$ is simply the target polarization vector.
For the spin-1 case $\bm S$ is defined in terms of polarization vectors
$\bm{e}^{m}$ as $\bm{S}=i{\bm{e}^m}^*{\times}\bm{e}^m$, where $m=0,\pm 1$
is the spin projection along the axis of quantization.

The nuclear spectral function can in general be calculated by
inserting a complete set of intermediate states and computing the
resulting transition matrix elements between the ground and
intermediate states.
For the deuterium nucleus, the intermediate states are exhausted by
a single proton or neutron, and the spectral function is expressed
entirely in terms of the deuteron wave function.
However, already for $A=3$ nuclei the calculation of the spectral
function is considerably more complicated \cite{SS,Italy}.

The general spin structure of the spectral function can be obtained
by expanding the matrix $\mathcal P^\tau_{\sigma\sigma'}$ in terms
of the Pauli spin matrices and applying constraints from parity and
time reversal invariance.
One can then write the spectral function in the general form
(for both the proton and neutron contributions) \cite{SS}:
\begin{equation}
\label{eq:fdef}
\spfn{\tau}{\bm{S}} =  \frac12
\left(
   f_0^\tau\ \bm{I}
 + f_1^\tau\,\bm{\sigma}\cdot\bm{S}
 + f_2^\tau\, T_{ij}\, S_i\, \sigma_j    
\right)\ ,
\end{equation}
where $\bm I$ is the unity matrix and
$T_{ij} = \widehat p_i\ \widehat p_j - \tfrac13\ \delta_{ij}$
is a traceless symmetric tensor with $\widehat p_i=p_i/|\bm{p}|$
for any component $i$ of the momentum
(the sum over repeated indices is implied).
%

Since the spectral function is hermitian, the coefficients $f_i^\tau$
in \eq{eq:fdef} are real functions of the energy $\ceps$ and momentum
$\bm{p}$.
The function $f_0^\tau$ gives the spin averaged spectral function,
and is normalized to the number of protons ($\tau=p$) or neutrons
($\tau=n$) in the nucleus:
\begin{equation}
\int [\ud p]\ \tr\left[\spfn{p(n)}{S} \right]
= \int [\ud p]\, f_0^{p(n)} = Z\ (A-Z)\ .
\label{eq:f0norm}
\end{equation}
The functions $f_1^\tau$ and $f_2^\tau$ describe the distribution
of nuclear spin amongst the nucleons.
The integrated functions $f_1^\tau$ and $f_2^\tau$ determine the
average nucleon polarization in the target and the average tensor
polarization, respectively:
\begin{align}
\left\langle\sigma_z\right\rangle^\tau
&= \intp\, f_1^\tau\ ,				\label{eq:f1norm} \\
\left\langle T_{zi}\ \sigma_i \right\rangle^\tau
&= \frac29 \intp\, f_2^\tau\ ,			\label{eq:f2norm}
\end{align}
where we take the nuclear spin vector to lie along the $z$-axis.

\vspace*{1cm}
{\em 2 c. Master formula}
\vspace*{0.5cm}

Projecting from the hadronic tensor (\ref{WA:NR}) the appropriate
structure functions, we obtain our ``master formula'' for the $g_1^A$
and $g_2^A$ structure functions of the nucleus:
\begin{equation}
xg_a^A(x,Q^2) = 
\int[\ud p]\,  
           D_{ab}^\tau(\ceps,\bm{p},\gamma)\,
           x' g_b^\tau(x',Q^2,p^2)\ ,
\label{eq:master}
\end{equation}
where $a,b=1,2$, and again summation over repeated indices is implied.
In Eq.~(\ref{eq:master})
$x' = Q^2/2p\cdot q = x/(1 + (\varepsilon+\gamma p_z)/M)$
is the Bjorken variable for the bound nucleon, with
$x=Q^2/2 M q_0$, and
$\gamma = |\bm{q}|/q_0 = \sqrt{1 + 4 M^2 x^2/Q^2}$
takes into account finite-$Q^2$ kinematics.
The spin-dependent nucleon momentum distribution functions $D_{ab}$
can be written in terms of the coefficient functions $f_i$ (dropping
isospin labels) as:
\begin{subequations}\label{eq:Dab}
\begin{align}
D_{11} =&
     f_1
  + \frac{3-\gamma^2}{6\gamma^2}\left(3\widehat p_z^2-1\right) f_2
  + \frac{v\widehat p_z}{\gamma}\left(f_1 + \frac{2}{3} f_2\right)
\notag
\\
&
  {} + v^2\frac{(3-\gamma^2)\widehat p_z^2-1-\gamma^2}{12\gamma^2}(3f_1-f_2),
\label{eq:D11}\\
D_{12} =&
     (\gamma^2-1)
\left[
      - \frac{3\widehat p_z^2-1}{2\gamma^2}\, f_2
  + \frac{v\widehat p_z}{\gamma}
    \left(f_1+\left(\frac{3}{2}\widehat p_z^2-\frac{5}{6}\right)\,f_2
    \right)
\right.
\notag
\\ &
\left.
    {} - v^2 \left(
  \frac{1+\widehat p_z^2(4\gamma^2-3)}{4\gamma^2}\,f_1
+\frac{5+18\widehat{p}_z^4\gamma^2-5\widehat{p}_z^2(3+2\gamma^2)}
      {12\gamma^2}\,f_2
            \right)
\right],
\label{eq:D12}\\
D_{21} =&
     -\frac{3\widehat{p}_z^2-1}{2\gamma^2} f_2
    -\frac{v\widehat{p}_z}{\gamma}\left(f_1 + \frac{2}{3} f_2 \right)
  -v^2 \frac{3\widehat{p}_z^2-1}{12\gamma^2}\left(3f_1-f_2 \right),
\label{eq:D21}\\
D_{22} =&
     f_1 + \frac{2\gamma^2-3}{6\gamma^2}\left(3\widehat{p}_z^2-1\right)f_2
\notag
\\
&
   {} + \frac{v\widehat{p}_z}{\gamma}
     \left[
     (1-\gamma^2) f_1
      + \left( -\frac{5}{6} + \frac{\gamma^2}{3}
               + \widehat{p}_z^2(\frac{3}{2} - \gamma^2
        \right) f_2
     \right]
\notag 
\\
&
\qquad\quad
{} + v^2\left[
 \frac{\widehat{p}_z^2(3-6\gamma^2+4\gamma^4)-1-2\gamma^2}{4\gamma^2}\,f_1
\right.
\notag
\\
&
\left.\qquad\qquad\quad
  {} + \frac{5-2\gamma^2(1+3\widehat{p}_z^2)+4\widehat{p}_z^2\gamma^4}
       {12\gamma^2}(3\widehat{p}_z^2-1) f_2
      \right]\ ,
\label{eq:D22}
\end{align}
\end{subequations}
where the velocity parameter $v = |\bm{p}|/M$.

The derivation of \eqs{eq:Dab} makes use of the fact that characteristic
values of $v$ are small, and one can expand the kinematical factors in a
series in $v$ keeping terms up to ${\cal O}(v^2)$.
The functions $D_{ab}$ have also been averaged over the polar angle
of the nucleon momentum in the transverse plane $(p_x,p_y)$ using the
independence of $f_{1,2}$ of the directions of the nucleon momentum
and the independence of the Bjorken variable $x'$ of $p_x$ and $p_y$.

In general the structure functions $g_{1,2}^\tau(x',Q^2,p^2)$ of the
bound nucleon are functions of the invariant mass squared of the
nucleon, since $p^2 \neq M^2$.
This dependence is {\em a priori} unknown, but has been estimated
in various models \cite{MST,KPW,AKL,KP,KMPW}.
In the WBL, however, and especially for light nuclei such as deuterium,
the degree to which the nucleons are off-mass-shell is not large,
and we can assume that the off-shell functions can be approximated by
their on-shell values,
$g_{1,2}^\tau(x',Q^2,p^2) \approx g_{1,2}^\tau(x',Q^2)$.

Note that while the above derivation is valid in the WBL
($|\bm{p}|, |\varepsilon| \ll M$), Eq.~(\ref{eq:master}) holds
for arbitrary momentum transfer $q$.
An interesting feature of the distribution functions $D_{ab}$ is that
they depend on the momentum transfer through the dimensionless parameter
$\gamma$.
Furthermore, nuclear effects cause the off-diagonal distributions
$D_{12}$ and $D_{21}$ to be nonzero, which results in mixing of
the $g_1$ and $g_2$ structure functions in the convolution integral
(\ref{eq:master}).
In leading order in $v$ the functions $D_{12}$ and $D_{21}$
are driven by tensor distribution $(3\hat p^2_z{-}1)f_2$.
In the limit of high $Q^2$, the parameter $\gamma \to 1$ and the
distributions (\ref{eq:Dab}) simplify considerably.
In particular, $D_{12} \to 0$ and the convolution formula for $g_1$
becomes diagonal ({\em i.e.} there are no contributions from
$g_2^\tau$ to $g_1^A$).
However, mixing in $g_2^A$ persists even in this limit \cite{KMPW}.

A similar mixing of the nucleon structure functions $g_1^\tau$ and
$g_2^\tau$ in the $g_1^A$ nuclear structure function was observed
in Refs.~\cite{SS,Italy,ScopHe3}.
However, the distribution functions in \eq{eq:Dab} are different
from the corresponding results in Refs.~\cite{SS,Italy} in the
higher order terms in $v$ and $\gamma^2-1$.
As discussed at the end of Sec.~2a, these differences arise from
the different treatments of the impulse approximation here and in
Refs.~\cite{SS,Italy,ScopHe3}.

The fact that Eq.~(\ref{eq:master}) can be applied to both the inelastic
and quasi-elastic scattering, at any $Q^2$, allows us calculate nuclear
structure functions in the DIS region (at large $Q^2$ and $W$), as well
as at lower $Q^2$, where the low-$W$ resonance region plays a more
prominent role.
In this context we note that the dependence of the effective nuclear
distributions \eq{eq:Dab} on $Q^2$ enters through the dimensionless
parameter $\gamma$, a feature which allows a simple parameterization
of the effective distributions over the full range of kinematics
\cite{KM}.
In the next section we apply the formal results presented here to the
specific case of the deuteron.

\vspace*{1cm}
{\bf 3. Deuteron spin structure functions}
\vspace*{0.5cm}

For the case of lepton scattering from a deuteron, the functions $f_i$ in
Eq.~(\ref{eq:fdef}) can be written in terms of the deuteron wave functions
as (see also Appendix~B of Ref.~\cite{KMPW}):
\begin{subequations}
\label{eq:f}
\begin{align}
f_0 &= 	4\pi^3\
	(\psi_0^2 + \psi_2^2)\
	\delta(\varepsilon - \epsilon_D + \bm{p}^2/2M)\ , \\
f_1 &= 	4\pi^3\
	(\psi_0^2 - \psi_2^2/2)\
	\delta(\varepsilon - \epsilon_D + \bm{p}^2/2M)\ , \\
f_2 &=  4\pi^3\
	\frac32 (\psi_2^2 - \sqrt2 \psi_0 \psi_2)\
	\delta(\varepsilon - \epsilon_D + \bm{p}^2/2M)\ ,
\end{align}
\end{subequations}
where $\epsilon_D = -2.2$ MeV is the deuteron binding energy,
and $\psi_0$ and $\psi_2$ are the $S$- and $D$-state momentum
space wave functions, respectively, normalized such that:
\begin{equation}
\label{eq:wfnorm}
\int_0^\infty\!\! \ud p\, \bm{p}^2
\left( \psi_0^2(p) + \psi_2^2(p) \right) = 1\ .
\end{equation}
Since the deuteron is an isoscalar nucleus, Eqs.~(\ref{eq:f}) hold
for both the proton and neutron distributions.
The average nucleon polarization and tensor polarization in the
polarized deuteron can then be expressed as:
\begin{align}
\langle\sigma_z\rangle &= 1-\frac32 P_D\ ,\\
\langle T_{zi}\ \sigma_i\rangle &= \frac13 (P_D-\sqrt2 P_{SD})\ ,
\end{align}
where
$P_D=\int \ud p\, \bm{p}^2 \psi_2^2(p)$ is the $D$-state probability,
and
$P_{SD}=\int \ud p\, \bm{p}^2 \psi_0(p)\psi_2(p)$ is the $S$-$D$
interference in the deuteron.
For the deuteron wave function calculated from the Paris potential
\cite{Paris} one has $P_D = 5.8\%$ and $P_{SD} = 9.4\%$, while for
the Bonn potential \cite{Bonn} $P_D = 4.3\%$ and $P_{SD} = 10.1\%$.
Note that in the $\gamma \to 1$ limit, the distributions $D_{ab}$ in
Eqs.~(\ref{eq:Dab}) for the deuteron are equivalent to those in
Ref.~\cite{KMPW}, where the functions $f_i$ were effectively defined
including a factor $(1-\bm{p}^2/2M^2)$, which in our notation is now
contained inside the distributions $D_{ab}$.

As discussed in Sec.~2 above, at finite $Q^2$ the deuteron structure
functions $g_{1,2}^d$ receive contributions from both the isoscalar
$g_1^N$ and $g_2^N$ structure functions of the nucleon ($N = p+n$)
individually.
In particular, while the contribution from $g_2^N$ to $g_1^d$ vanishes
in the Bjorken limit, it is non-zero at finite $Q^2$.
To illustrate the relative importance of the $g_{1,2}^N$ contributions
to the deuteron structure functions, we can compare the individual
momentum distributions $D_{ab}$ for the diagonal and off-diagonal
terms, and also as a function of $\gamma$.

In the simplest convolution model the nuclear structure functions in
Eq.~(\ref{eq:master}) are written as convolutions of the nucleon
structure functions and effective nucleon {\em light-cone} momentum
distributions:
\begin{eqnarray}
g_a^d(x,Q^2)
&=& 
    \int_x \frac{\ud y}{y}\ \widetilde D_{ab}(y,\gamma)\
            g_b^N\left(\frac{x}{y},Q^2\right)\ ,
\label{eq:gadLC}
\end{eqnarray}
where $y = (p_0 + \gamma p_z)/M
= (1 + (\varepsilon+\gamma p_z)/M) = x/x'$.
In the Bjorken limit ($\gamma\to 1$) the variable $y$ is the light-cone
fraction of the deuteron carried by the interacting nucleon.
The effective light-cone momentum distributions $\widetilde D_{ab}$ are
obtained by integrating the functions $D_{ab}$ in Eq.~(\ref{eq:Dab}):
\begin{eqnarray}
\widetilde D_{ab}(y,\gamma) &=& \int [\ud p]\
    D_{ab}(\ceps,\bm{p},\gamma)\
\delta \left(y-1-\frac{\ceps +\gamma p_z}{M} \right).
\end{eqnarray}

\begin{sidewaysfigure}[p]
\begin{center}
\epsfig{file=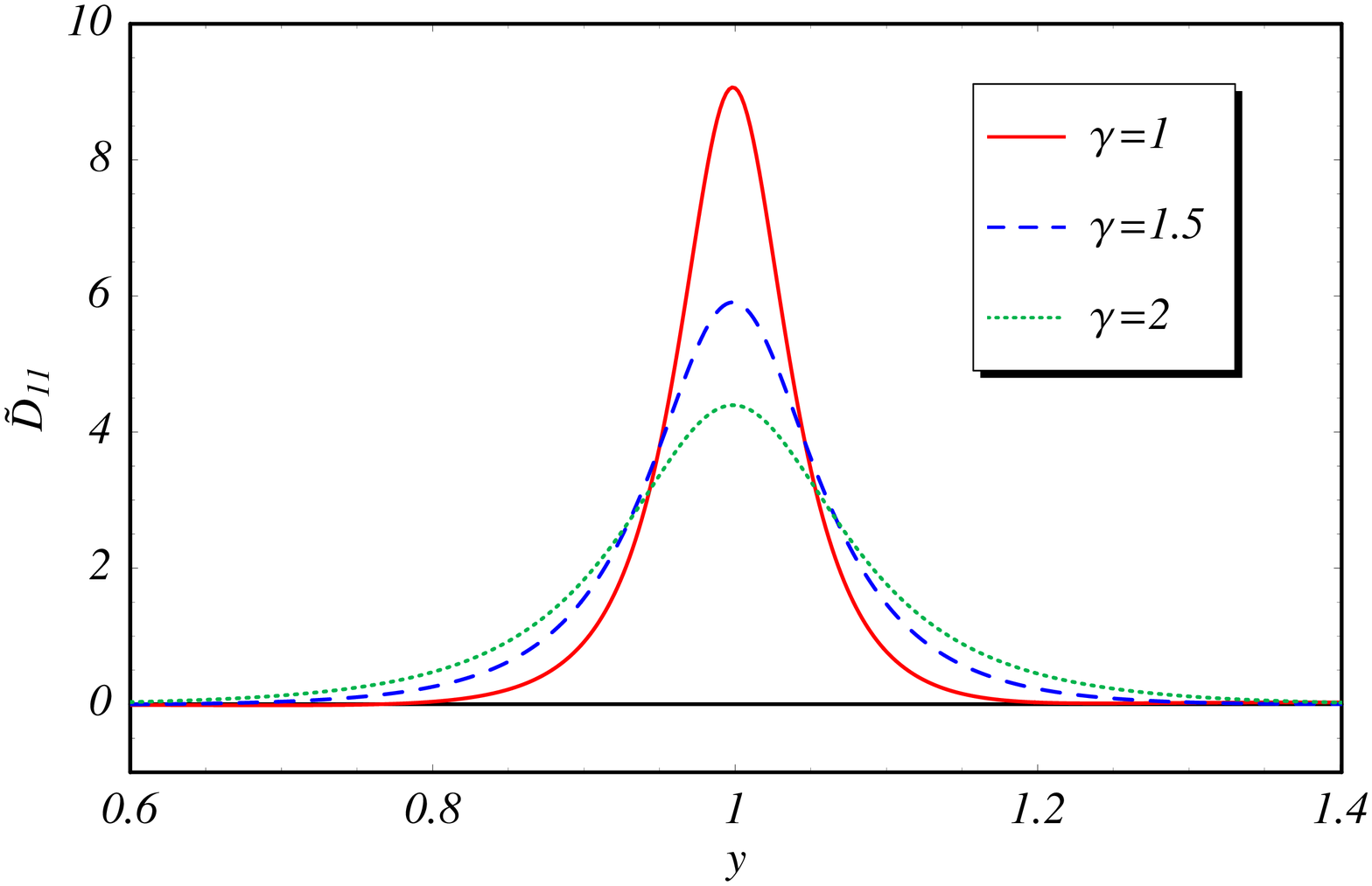,width=12.0cm}%
\epsfig{file=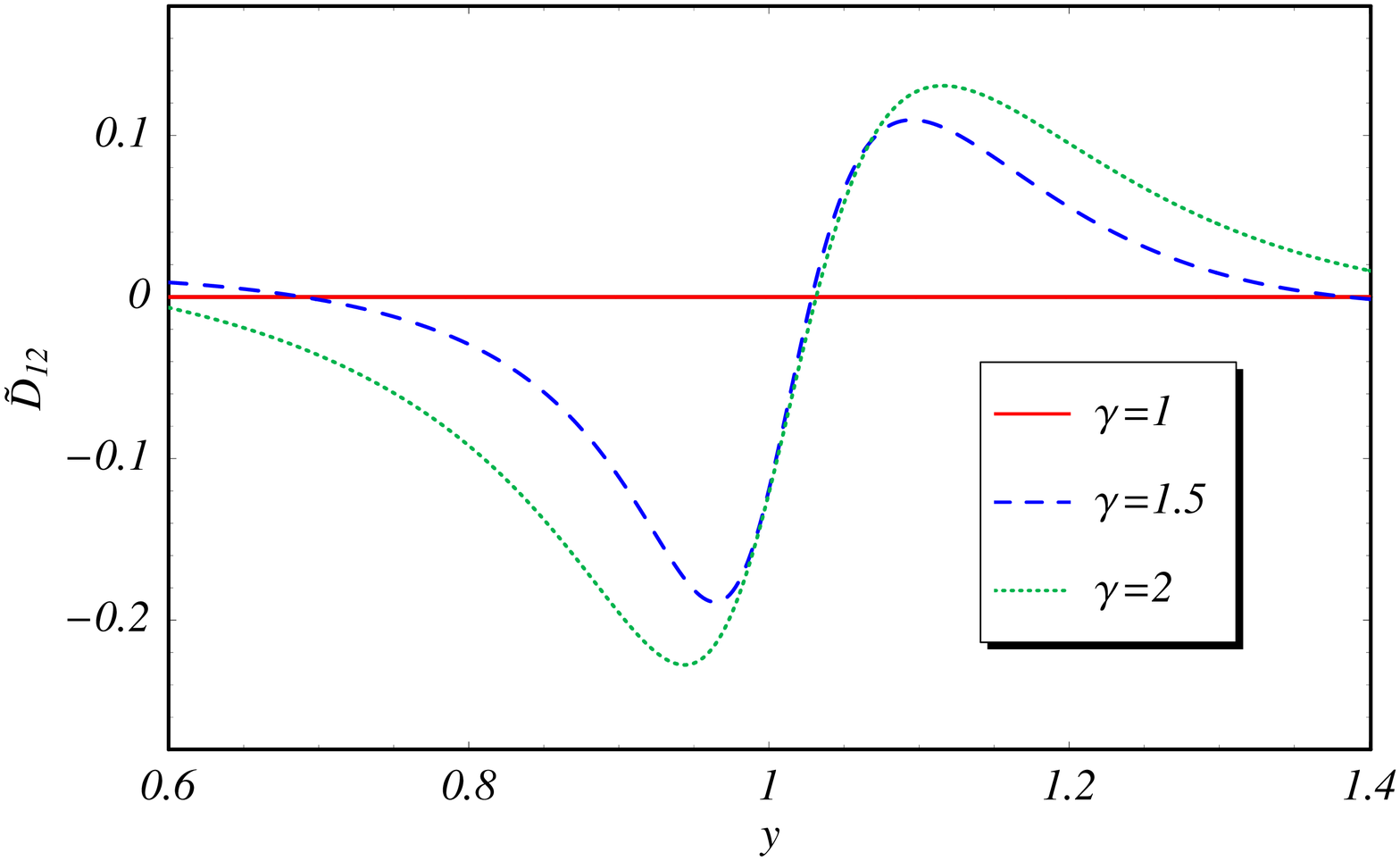,width=12.0cm}
\epsfig{file=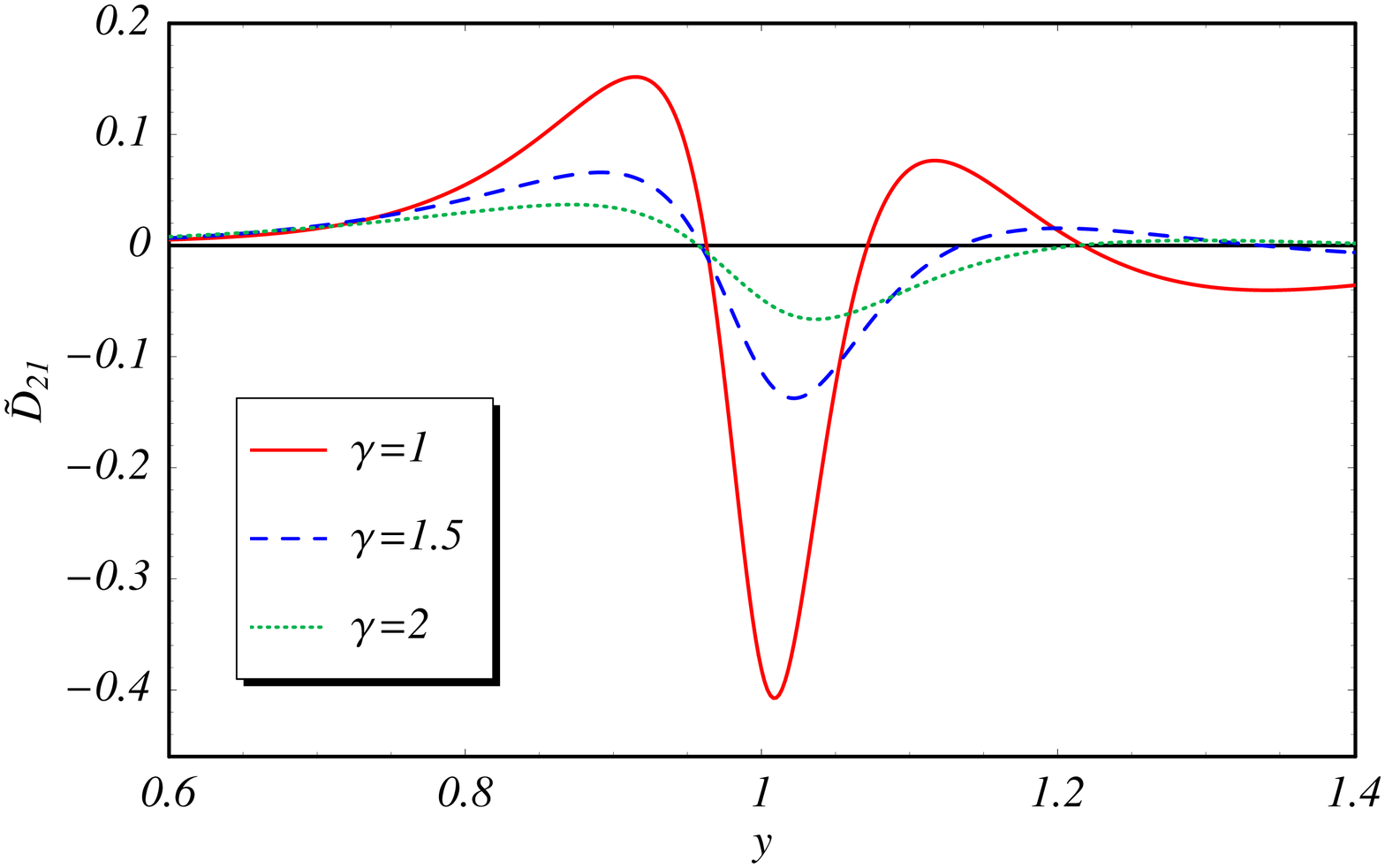,width=12.0cm}%
\epsfig{file=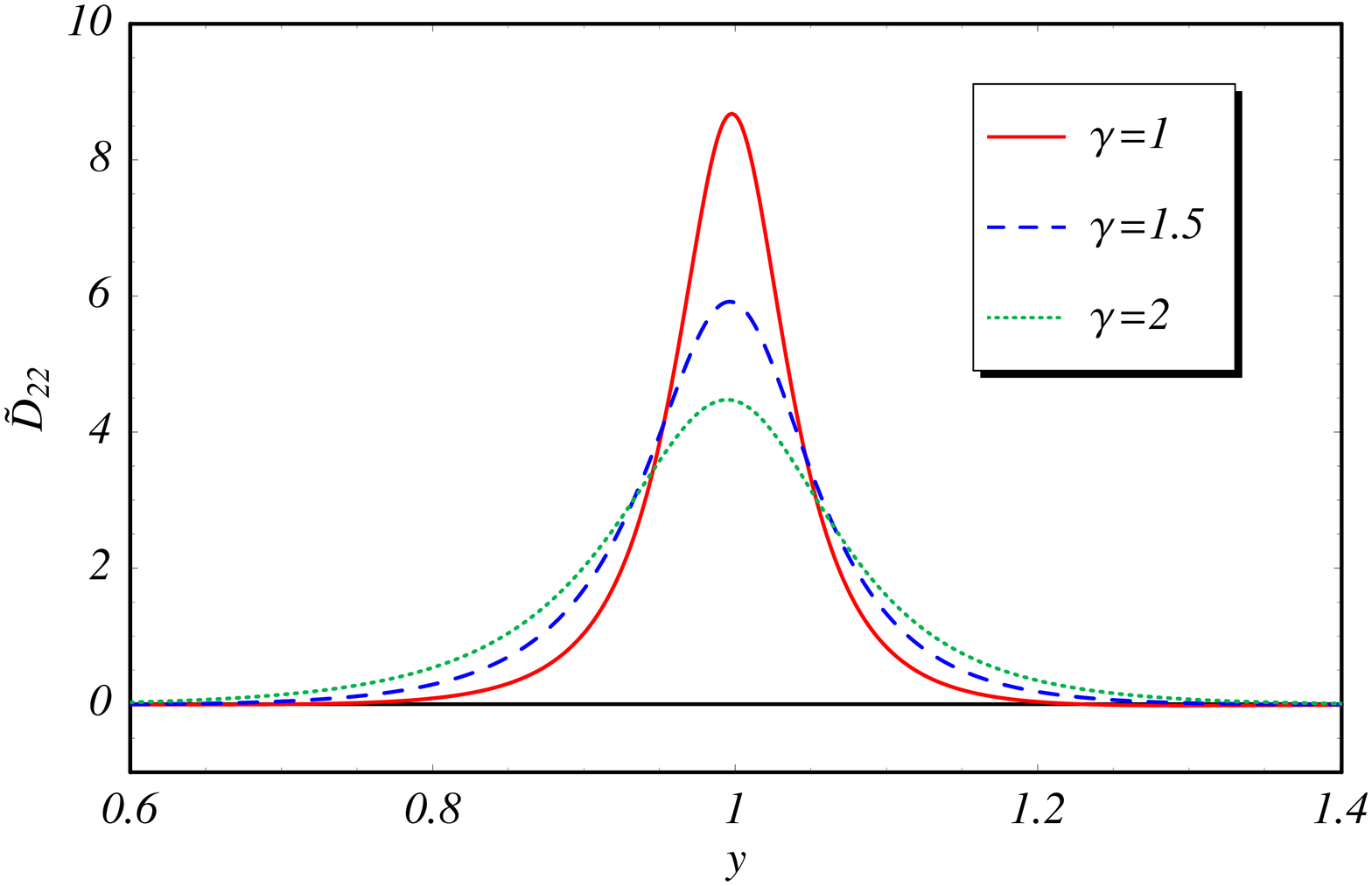,width=12.0cm}
\caption{Effective nucleon light-cone momentum distribution functions
	$\widetilde D_{ab}(y,\gamma)$ for $\gamma=1$ (Bjorken limit),
	1.5 and 2.}
\label{fig:Dab}
\end{center}
\end{sidewaysfigure}

In Fig.~\ref{fig:Dab} we show the nucleon light-cone momentum
distributions $\widetilde D_{ab}(y,\gamma)$ for several values of
$\gamma$.
The results for $\gamma = 1$ correspond to the Bjorken limit
distributions.
Note that in this limit the function $\widetilde D_{12}$ vanishes.
The diagonal functions $\widetilde D_{11}$ and $\widetilde D_{22}$
are significantly larger than the off-diagonal functions, but
decrease in magnitude for larger $\gamma$.
On the other hand, the distribution $\widetilde D_{12}$ becomes
larger with increasing $\gamma$, with its magnitude reaching
$\sim 5\%$ that of $\widetilde D_{11}$ at $\gamma=2$.

In Ref.~\cite{ScopD} the resonance region was studied using a
finite-$Q^2$ convolution formula for $g_1^d$ similar to that in
Eq.~(\ref{eq:gadLC}).
However, the non-diagonal term arising from $g_2^N$ was not present.
We find that this term, although small, does make a non-zero
contribution at finite values of $Q^2$.

In the next section we use Eq.~(\ref{eq:master}) to evaluate the
effects of the smearing of the nucleon structure functions by the
nucleon momentum distributions, and compare the calculated deuteron
structure functions with the input nucleon structure functions in
the resonance and deep inelastic regions.

\vspace*{1cm}
{\bf 4. Nuclear effects}
\vspace*{0.5cm}

In this section we present results for the $g_1^d$ and $g_2^d$
structure functions of the deuteron, and compare these with the
free nucleon functions.
We focus in particular on the resonance region, with $W \lesssim 2$~GeV,
which has to date received little attention.
The need to understand nuclear effects in the deuteron $g_1^d$ and
$g_2^d$ structure functions has arisen partly in response to the recent
high-precision data on the spin dependent deuteron structure functions
from Jefferson Lab \cite{EG1b,EG4,RSS}.

For the nucleon $g_1^\tau$ and $g_2^\tau$ structure functions, in this
analysis we consider the parameterizations from the MAID unitary isobar
model for $(e,e'p)$ reactions \cite{N:MAID}, from Simula {\em et al.}
\cite{N:Simula}, and from EG1 in CLAS at Jefferson Lab \cite{N:Kuhn}.
These parameterizations encompass the resonance as well as the deep
inelastic regions.
In the case of the MAID model, however, only a few selected final states
are considered, so this parameterization is expected to underestimate
the high-$W$ (or low-$x$) region.

\begin{figure}[ht]
\includegraphics[width=13cm]{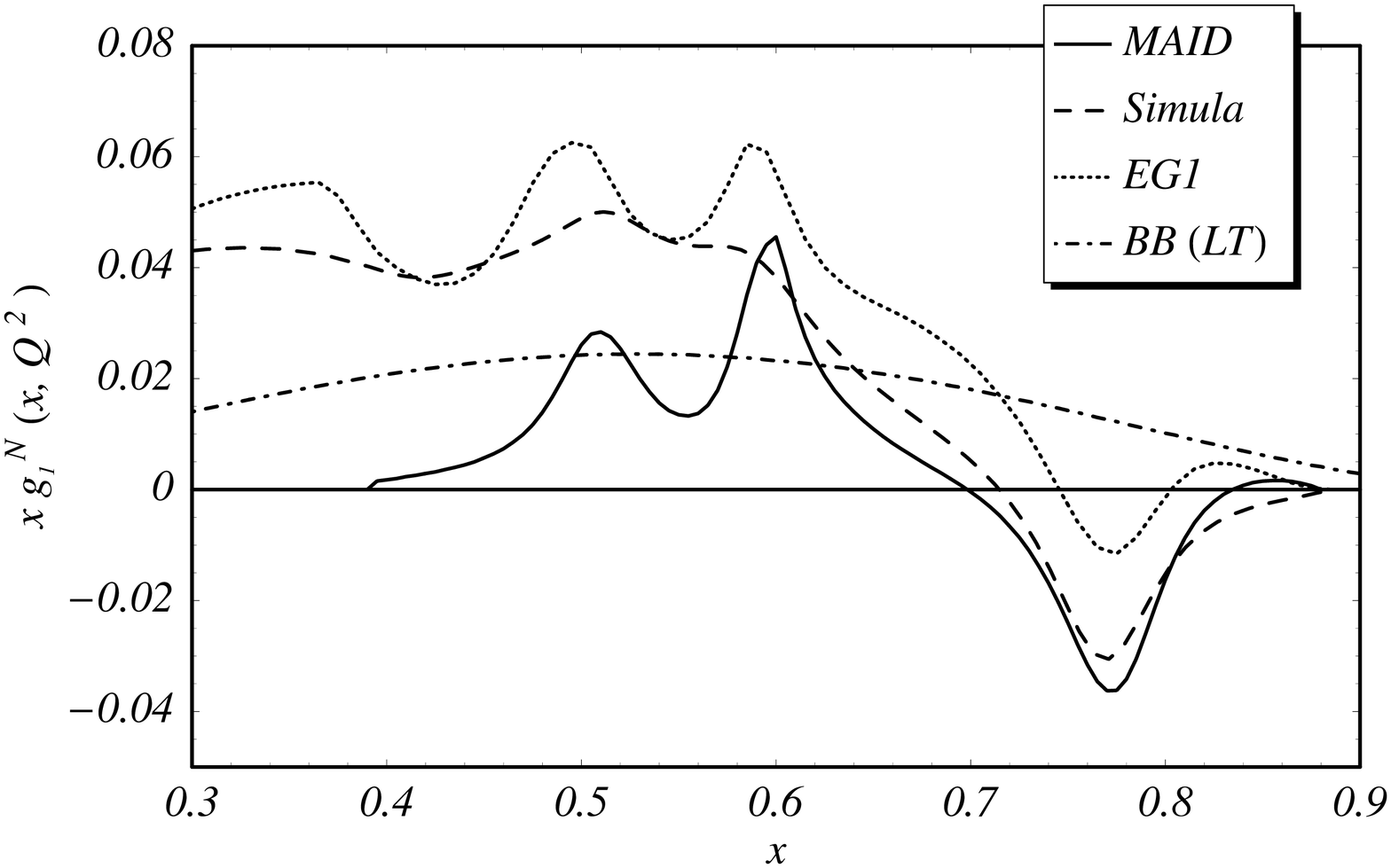}
\caption{Spin-dependent $x g_1^N$ structure function of an isoscalar
	nucleon, from the fits of Refs.~\cite{N:MAID} (solid),
	\cite{N:Simula} (dashed), \cite{N:Kuhn} (dotted), and the
	leading twist fit of Ref.~\cite{N:BB} (dot-dashed), at
	$Q^2 = 2$~GeV$^2$.}
\label{fig:g1N}
\end{figure}

\begin{figure}[hb]
\includegraphics[width=13cm]{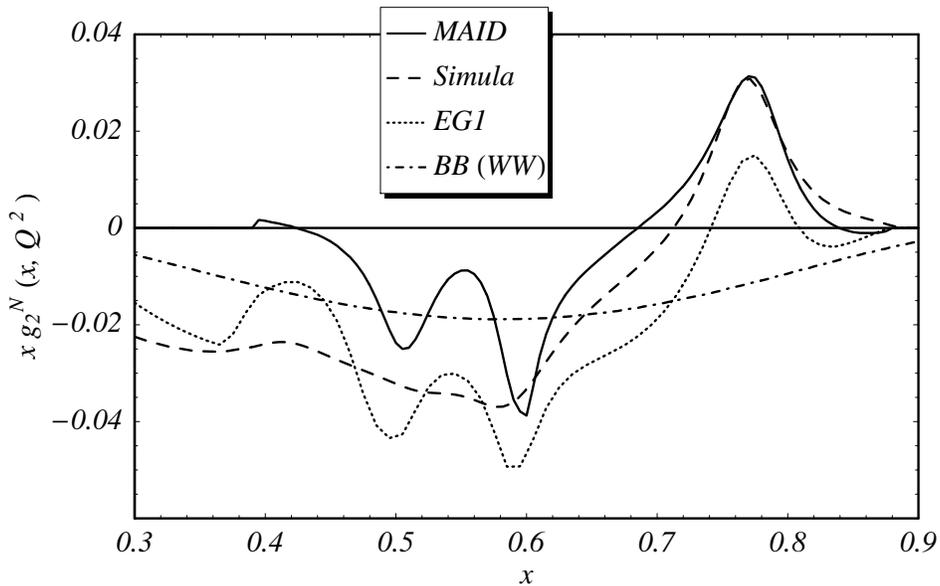}
\caption{Spin-dependent $x g_2$ structure function of an isoscalar
	nucleon, from the fits of Refs.~\cite{N:MAID} (solid),
	\cite{N:Simula} (dashed), \cite{N:Kuhn} (dotted), and the
	Wandzura-Wilczek approximation to $g_2$ using the leading
	twist $g_1$ fit of Ref.~\cite{N:BB} (dot-dashed),
	at $Q^2 = 2$~GeV$^2$.}
\label{fig:g2N}
\end{figure}

In Figs.~\ref{fig:g1N} and \ref{fig:g2N} we show the input $xg_1^N$
and $xg_2^N$ structure functions, respectively, for an isoscalar
nucleon $(N=p+n)$, at a sample $Q^2$ value, $Q^2 = 2$~GeV$^2$.
At this $Q^2$ the most prominent structure at large $x$ in both the
$g_1^N$ and $g_2^N$ structure functions is the peak associated with
the $P_{33}$ $\Delta$ resonance, which is negative for $g_1^N$ and
positive for $g_2^N$.
The three models give qualitatively similar results here, although
quantitatively there are some differences.
At lower $x$ the second and third resonance regions are also prominent,
and here the MAID fit is smaller in magnitude, as expected, given that
it is constructed primarily to describe low-$W$ data.

For comparison in Fig.~\ref{fig:g1N} we also show the leading twist
(``BB'') parameterization of $xg_1^N$ from Ref.~\cite{N:BB}, which
is smooth and does not contain any resonance structure.
It is interesting to observe that the leading twist fit appears to
go through the average of the resonances (with the exception of the
$\Delta$ resonance) for the MAID fit \cite{N:MAID}, reminiscent of
the Bloom-Gilman duality between resonance and DIS structure
functions \cite{MEK}.
The other resonance parameterizations \cite{N:Simula,N:Kuhn} are on
average larger in magnitude than the leading twist curve, which may
reflect the presence of the nonresonant background that is included
in these fits.

Using the BB parameterization \cite{N:BB} we also calculate the
leading twist Wandzura-Wilczek (WW) approximation to $g_2$,
$g_2^{\rm WW}(x,Q^2) = - g_1(x,Q^2) + \int_x^1 dy\ g_1(y,Q^2)/y$,
which is shown in Fig.~\ref{fig:g2N}.
This again displays very different behavior, both in magnitude and sign,
compared with the $g_2^N$ resonance parameterizations at large $x$.
As with the $g_1^N$ comparison, the leading twist curve appears to
average the second and third resonance regions at intermediate $x$ as
described by the MAID \cite{N:MAID} fit, but is smaller in magnitude
here compared with the other fits \cite{N:Simula,N:Kuhn}.

\begin{figure}[ht]
\includegraphics[width=13cm]{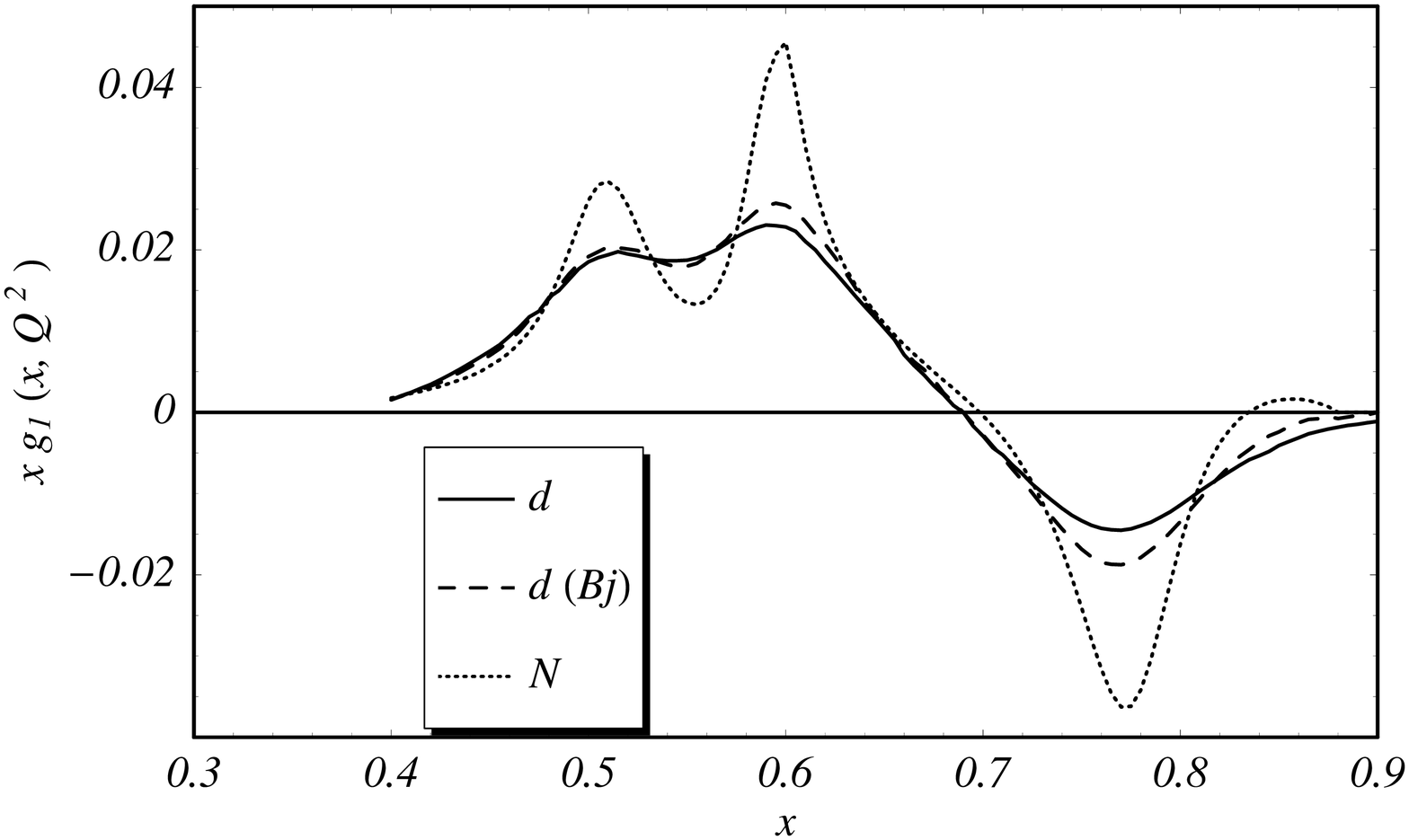}
\caption{Spin-dependent $x g_1$ structure function of the deuteron
	evaluated at finite-$Q^2$ (solid) and Bjorken limit (dashed)
	kinematics, compared with the nucleon (dotted) input, from
	Ref.~\cite{N:MAID} at $Q^2 = 2$~GeV$^2$.}
\label{fig:g1dM}
\end{figure}

\begin{figure}[hb]
\includegraphics[width=13cm]{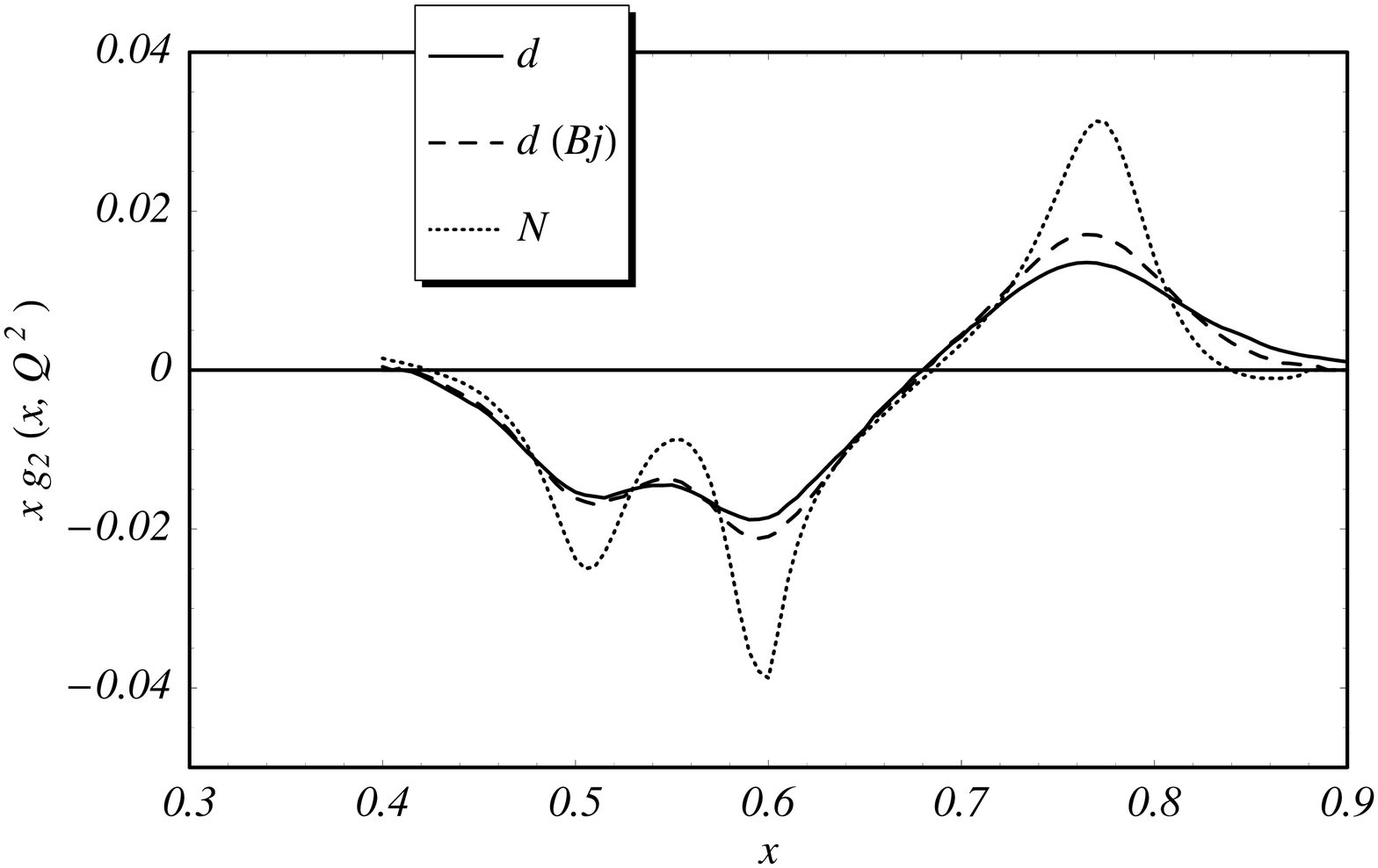}
\caption{Spin-dependent $x g_2$ structure function of the deuteron
	evaluated at finite-$Q^2$ (solid) and Bjorken limit (dashed)
	kinematics, compared with the nucleon (dotted) input, from
	Ref.~\cite{N:MAID} at $Q^2 = 2$~GeV$^2$.}
\label{fig:g2dM}
\end{figure}

With these nucleon structure function parameterizations, we can now
investigate nuclear effects in both the resonance and DIS regions.
For the resonance region ($W \lesssim 2$~GeV), we use the MAID
parameterization \cite{N:MAID} for $g_1^N$ and $g_2^N$.
The deuteron $xg_1^d$ structure function, calculated using
Eqs.~(\ref{eq:master}), (\ref{eq:D11}) and (\ref{eq:D12}) with the
Paris deuteron wave function, is shown in Fig.~\ref{fig:g1dM} at
$Q^2 = 2$~GeV$^2$ (solid curve).
Here the full calculation of $xg_1^d$ is compared with that using the
Bjorken limit in the momentum distributions $D_{11}$ and $D_{12}$
(dashed), and with the free nucleon structure functions (dotted).
Where the resonance structures are clearly evident in the nucleon
$g_1^N$ functions, they are significantly diluted in the deuteron
structure function.
At the peak of the $\Delta$ resonance, for example, the magnitude
of the deuteron $g_1^d$ is about half that of the nucleon $g_1^N$.
A similar effect is seen for the $g_2^d$ structure function shown
in Fig.~\ref{fig:g2dM}.

The differences between the full results for $g_{1,2}^d$ and the
Bjorken limit ($Q^2 \to \infty$) approximation for $D_{ab}$ are small
where the nucleon $g_{1,2}^N$ are smooth, but become more significant
in the vicinity of the resonance peaks.
At intermediate $x$, around the second and third resonance regions,
the full results are some 10--15\% smaller in magnitude than the
Bjorken limit structure functions.
At larger $x$ the enhanced smearing is even more dramatic, with the
full results being up to 25--30\% smaller at the $\Delta$ resonance
peak than those with the Bjorken limit smearing.
This behavior can be understood from the nucleon light-cone
distribution functions in Fig.~\ref{fig:Dab},
which decrease in magnitude with increasing $\gamma$.
At fixed $Q^2 = 2$~GeV$^2$, the third resonance region at
$x \approx 0.5$ corresponds to $\gamma \approx 1.5$, while for the
$\Delta$ peak at $x \approx 0.75$ one has $\gamma \approx 2$.
Therefore, at fixed $Q^2$, larger $x$ implies larger $\gamma$,
and hence stronger smearing effects in both $g_1^d$ and $g_2^d$.

\begin{figure}
\includegraphics[width=14cm]{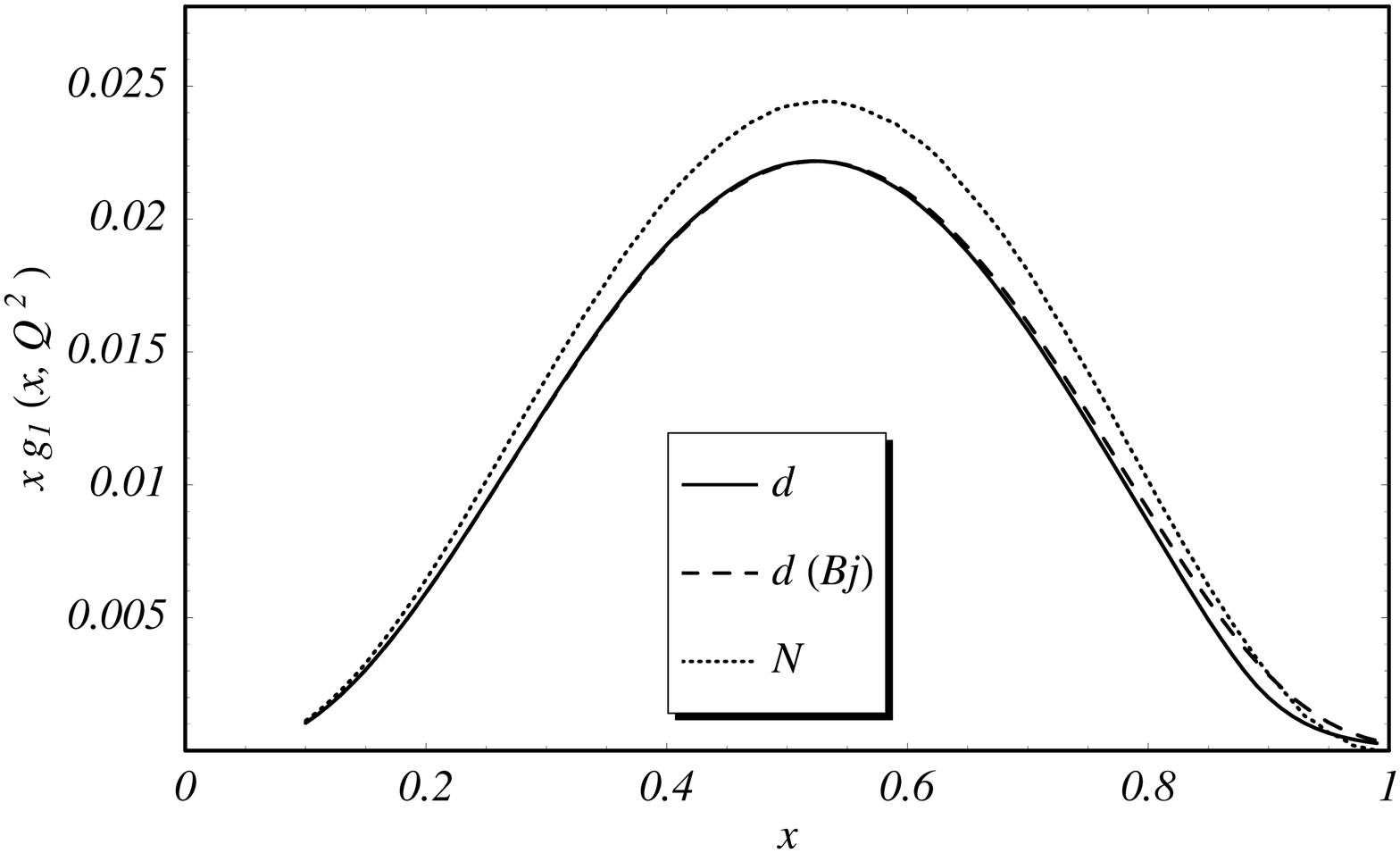}
\caption{Leading twist $x g_1$ structure function of the deuteron
	evaluated at finite-$Q^2$ (solid) and Bjorken limit (dashed)
	kinematics, compared with the nucleon (dotted) input from
	Ref.~\cite{N:BB} at $Q^2 = 2$~GeV$^2$.}
\label{fig:g1dB}
\end{figure}

\begin{figure}
\includegraphics[width=14cm]{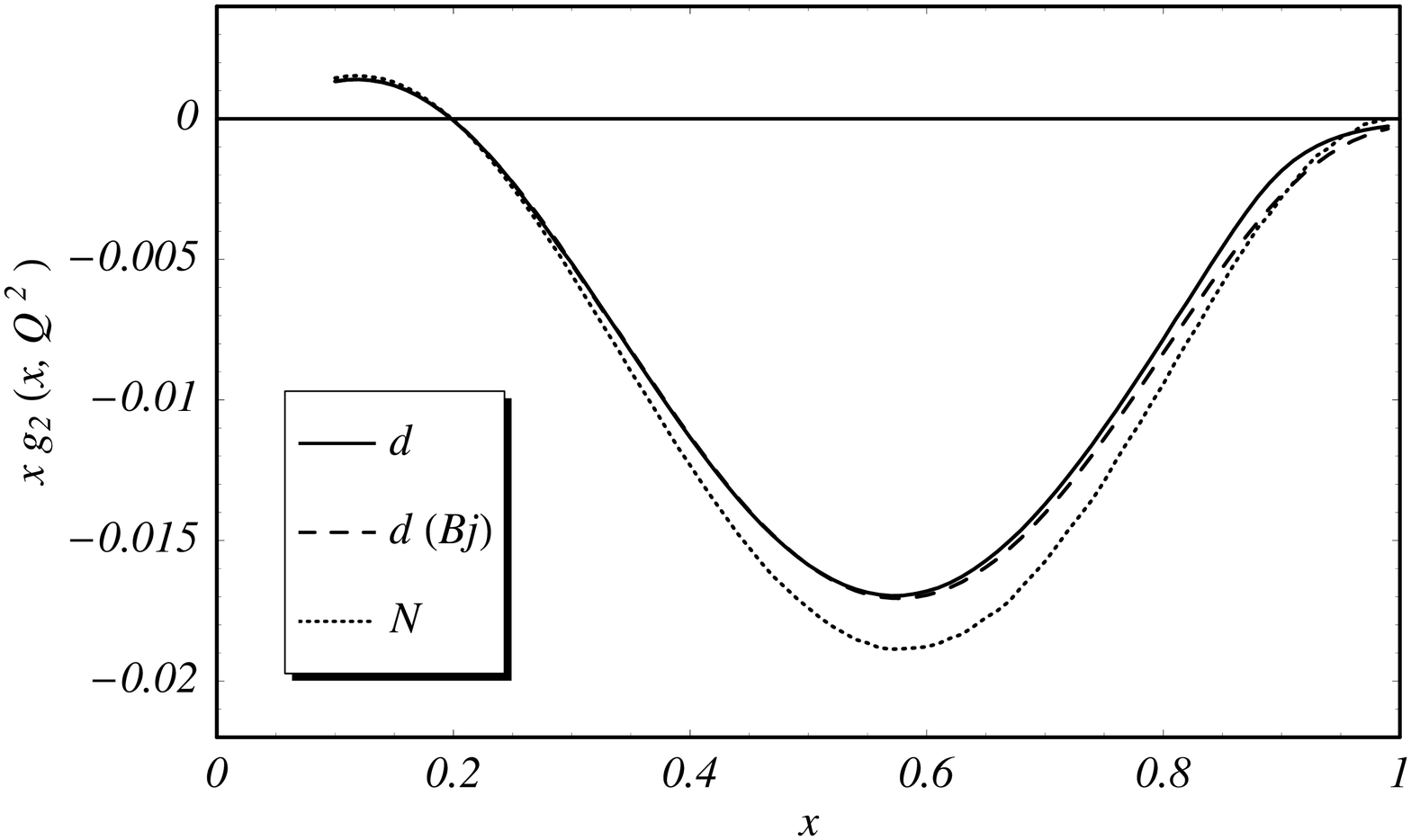}
\caption{Leading twist $x g_2$ structure function of the deuteron      
	evaluated at finite-$Q^2$ (solid) and Bjorken limit (dashed)
	kinematics, compared with the nucleon (dotted) input from   
	Ref.~\cite{N:BB} at $Q^2 = 2$~GeV$^2$.}
\label{fig:g2dB}
\end{figure}

The enhanced smearing at finite-$Q^2$ kinematics is also evident for
leading twist structure functions, although the effects here are less
dramatic.
In Figs.~\ref{fig:g1dB} and \ref{fig:g2dB} we show the deuteron
$xg_1^d$ and $xg_2^d$ structure functions, respectively, evaluated
using the leading twist ``BB'' parameterization of the nucleon
structure functions at $Q^2=2$~GeV$^2$.
The full calculation for $g_1^d$ and the Bjorken limit approximation
are similar, and both smaller in magnitude than $g_1^N$ for
$x \lesssim 0.8$.
Here to a good approximation the deuteron structure functions are
related to the nucleon structure functions by an $x$-independent
multiplicative factor, $g_{1,2}^d \approx (1-\frac32 P_D) g_{1,2}^N$.
At larger $x$, one sees the onset of a classic Fermi motion effect,
whereby the ratios $g_{1,2}^d/g_{1,2}^N$ increase in magnitude as
$x \to 1$.
In the region $0.8 \lesssim x \lesssim 1$, the finite-$Q^2$ results
are again smaller in magnitude than those for the Bjorken limit
kinematics, implying a stronger EMC effect at large $x$, for both
the $g_1^d$ and $g_2^d$ structure functions.

Although the shapes of the nucleon and deuteron structure functions
are very different, especially in the resonance region, when integrated
over $x$ the differences turn out to be remarkably small, particularly
for the lowest moment.
At large $Q^2$ the lowest moments of the nucleon and deuteron $g_{1,2}$
structure functions, $\Gamma_{1,2}(Q^2) = \int dx\ g_{1,2}(x,Q^2)$,
are to a very good approximation related by:
\begin{eqnarray}
\Gamma_{1,2}^d(Q^2)
&\approx& \left( 1 - \frac32 P_D \right) \Gamma_{1,2}^N(Q^2)\ ,
\end{eqnarray}
where for the Paris deuteron wave function the depolarization factor
relating the moments is $(1 - \tfrac32 P_D) \approx 0.91$.
Numerically, for the leading twist BB parameterization of the nucleon
structure functions \cite{N:BB}, using the Bjorken limit nucleon
momentum distributions we find
$\Gamma_1^{d\ (Bj)}/\Gamma_1^N = 0.91$ and
$\Gamma_2^{d\ (Bj)}/\Gamma_2^N = 0.91$
at $Q^2=2$~GeV$^2$.
Using the full expressions in Eqs.~(\ref{eq:master}) and (\ref{eq:Dab}),
on the other hand, we find the ratios
$\Gamma_1^d/\Gamma_1^N = 0.90$ and
$\Gamma_2^d/\Gamma_2^N = 0.90$.
Thus the finite-$Q^2$ kinematics slightly reduces the magnitude of
the deuteron moments, which is consistent with the observation of
the additional suppression at finite $Q^2$ at large $x$ in
Figs.~\ref{fig:g1dB} and \ref{fig:g2dB}.

The ratios of the deuteron to nucleon moments in the resonance region
remain similar to the leading twist ratios, even though the shapes of
the functions here are strongly $Q^2$ dependent and hence infused with
large higher twist contributions.
Specifically, for the MAID parameterization of $g_{1,2}^N$ the
ratios at finite-$Q^2$ kinematics turn out to be
$\Gamma_1^d/\Gamma_1^N = 0.91$ and
$\Gamma_2^d/\Gamma_2^N = 0.92$,
at the same $Q^2=2$~GeV$^2$.
Note that the structure functions here have been integrated from
$x_{\rm min} \approx 0.4$ up to $x_{\rm max} = x_{\rm thr}$, where
$x_{\rm thr} = Q^2/(W_{\rm thr}^2 - M^2 + Q^2)$ corresponds to the
kinematical pion production threshold, $W_{\rm thr} = M + m_\pi$.
The results using Bjorken limit kinematics differ from these only
in the third decimal point.

Overall, our results suggest that for the lowest moments of the $g_1$
and $g_2$ structure functions, the nuclear effects in the deuteron can
to very good accuracy be accounted for by applying the depolarization
correction, $(1-\tfrac32 P_D)$.
This will not be true, however, for higher moments, and certainly
this approximation will break down dramatically at large $x$,
for $x \geq 0.7-0.8$.

\vspace*{1cm}
{\bf 5. Conclusions}
\vspace*{0.5cm}

For most of the history of lepton--nucleus deep inelastic scattering,
the discussion of nuclear effects on structure functions has been
confined to analysis of high-energy data within theoretical frameworks
constructed to be valid in the limit of large $Q^2$ and $W^2$
($\gg M^2$).
Recent high-quality data at lower $Q^2$ and $W^2$, especially in the
transition region where nucleon resonances merge into the DIS continuum,
have revealed a richness of phenomena which had not previously been
appreciated because of the lack of precision in earlier data.
An accurate description and understanding of the new data clearly
demands comparable advances in the theoretical tools.

Our aim in this work has been to provide the framework for analyzing
modern high-precision data over the full range of kinematics where
they are available.
To this end we have derived relations between spin-dependent nuclear
and nucleon $g_1$ and $g_2$ structure functions in the weak binding
limit, which are valid at all $Q^2$, and in both the traditional DIS
region and the poorly explored nucleon resonance region.
As a consistency check, we verify that our results approach the
previously derived convolution formulas for the nuclear structure
functions at large $Q^2$ \cite{KMPW}.

We apply the formalism to the specific case of the deuteron, which is
often used as a source of neutron structure information, and compare
the size of the nuclear corrections calculated using exact kinematics
and using approximations applicable at large $Q^2$.
We find that significant smearing of the nucleon structure functions
occurs in regions where nucleon resonances are prominent, with the
exact results some 10--15\% smaller in magnitude than the Bjorken limit
structure functions around the second and third resonance regions.
The smearing is enhanced at larger $x$, where the full results are
up to 25--30\% smaller at the $\Delta$ resonance peak than those with
the Bjorken limit smearing.

The enhanced smearing at finite-$Q^2$ kinematics is also evident for
leading twist structure functions, although the effects here are less
dramatic.
At intermediate $x$ the deuteron structure functions are approximately
given by $g_{1,2}^d \approx (1-\frac32 P_D) g_{1,2}^N$.
However, at larger $x$, $0.8 \lesssim x \lesssim 1$, the finite-$Q^2$
results are again smaller in magnitude than those for the Bjorken limit
kinematics, implying a stronger EMC effect, for both the $g_1^d$ and
$g_2^d$ structure functions.

The ratios of the integrals of the deuteron and nucleon structure
functions are to a good approximation simply related by the
depolarization factor,
$\Gamma_{1,2}^d(Q^2) / \Gamma_{1,2}^N(Q^2) \approx (1 - \frac32 P_D)$,
even at low $Q^2$ ($Q^2 \sim 2$~GeV$^2$).
This holds for both the leading twist structure functions, and for
structure functions in the resonance region, where the shapes of the
functions are strongly $Q^2$ dependent, with deviations only of
${\cal O}(1\%)$.

In closing we should mention that our calculation is by no means
complete.
We have not considered, for example, possible modification of nucleon
properties (such as masses or widths) in the nuclear medium, which in
our framework amounts to neglecting nucleon dynamical off-mass-shell
effects in the nucleon structure functions.
Extensions in this direction can be carried out using existing models
for the off-shell extrapolation \cite{MST,MPT,KMPW,Cloet}; we have
chosen not to do so in this work in order to more clearly isolate the
effects associated with the finite-$Q^2$ kinematics.
Similarly, we have not considered here effects beyond the nuclear
impulse approximation, such as rescattering, meson exchange currents,
or final state interactions.
Finally, our formalism can be easily applied to other nuclei, such
as polarized $^3$He.
This will be the subject of a future publication \cite{KM}.

\begin{acknowledgments}

We thank S.~Choi, S.~Kuhn and G.~Salme for helpful communications.
W.M. is supported by the DOE contract No. DE-AC05-06OR23177, under
which Jefferson Science Associates, LLC operates Jefferson Lab.
S.K. is partially supported by the Russian Foundation for Basic
Research, project No. 06-02-16659 and 06-02-16353.

\end{acknowledgments}


\newpage

\vspace{-2cm}

\vspace{-2cm}

\vspace{-2cm}

\end{document}